\documentclass[conference]{IEEEtran}
\IEEEoverridecommandlockouts
\usepackage{cite}
\usepackage[acronym]{glossaries}
\usepackage{amsmath,amssymb,amsfonts}
\usepackage{algorithmic}
\usepackage{graphicx}
\usepackage{textcomp}
\usepackage{xcolor}
\usepackage{tabularx}
\usepackage{adjustbox}
\usepackage{todonotes}
\usepackage{balance}
\def\BibTeX{{\rm B\kern-.05em{\sc i\kern-.025em b}\kern-.08em
    T\kern-.1667em\lower.7ex\hbox{E}\kern-.125emX}}

\usepackage{cleveref}
\usepackage{enumitem}
\usepackage[all=normal,charwidths=tight,leading=tight]{savetrees}

\setenumerate{leftmargin=*}

\newacronym{JCAS}{JCAS}{Joint Communication and Sensing}
\newacronym{ICAS}{ICAS}{Integrated Communication and Sensing}
\newacronym{RAN}{RAN}{Radio Access Network}
\newacronym{SA}{SA}{standalone}
\newacronym{AI}{AI}{Artificial Intelligence}
\newacronym{MEC}{MEC}{Mobile Edge Computing}
\newacronym{ML}{ML}{Machine Learning}
\newacronym{CV2X}{C-V2X}{Cellular Vehicle-to-Everything}
\newacronym{V2X}{V2X}{Vehicle-to-Everything}
\newacronym{U2X}{U2X}{UAV-to-Everything}
\newacronym{UAV}{UAV}{Unmanned Aerial Vehicles}
\newacronym{V2V}{V2V}{Vehicle-to-Vehicle}
\newacronym{E2E}{E2E}{End-to-End}
\newacronym{QoS}{QoS}{Quality-of-Service}
\newacronym{CAM}{CAM}{Cooperative Awareness Message}
\newacronym{LiDAR}{LiDAR}{Light Detection and Ranging}
\newacronym{SAR}{SAR}{Synthetic Aperture Radar}
\newacronym{ITS}{ITS}{Intelligent Transportation Systems}
\newacronym{DT}{DT}{Digital Twin}
\newacronym{TC}{TC}{Tech Case}
\newacronym{RSU}{RSU}{Road Side Unit}
\newacronym{CL}{CL}{Convergence Level}
\newacronym{GPS}{GPS}{Global Positioning System}
\newacronym{SLAM}{SLAM}{Simultaneous Localization and Mapping}
\newacronym{GDPR}{GDPR}{General Data Protection Regulation}
\newacronym{PKI}{PKI}{Public-key Infrastructure}
\newacronym{mmW}{mmWave}{Millimeter wave}
\newacronym{BMBF}{BMBF}{Federal Ministry of Education and Research}
\newacronym{5GC}{5GC}{5G Core}
\newacronym{ORAN}{O-RAN}{Open Radio Access Networks}

\newacronym{PHY}{PHY}{Physical}
\begin{document}

\title{
Enabling Mobility-Oriented JCAS in 6G Networks: An Architecture Proposal\\
%Using Joint Communication and Sensing in Mobile Applications: A 6G Architecture\\
%Towards an inclusion of mobility-oriented Joint Communication and Sensing for the 6G Architecture
%Enabling mobility-oriented Joint Communication and Sensing in 6G networks: An architecture proposal
%Enabling mobility-oriented Joint Communication and Sensing Architecture for the 6G Communication System
}

\author{\IEEEauthorblockN{Philipp Rosemann\IEEEauthorrefmark{1}, Sanket Partani\IEEEauthorrefmark{1}, Marc Miranda\IEEEauthorrefmark{2}, Jannik Mähn\IEEEauthorrefmark{3}, Michael Karrenbauer\IEEEauthorrefmark{1}, William Meli\IEEEauthorrefmark{4},\\
Rodrigo Hernang\'{o}mez\IEEEauthorrefmark{6}, Maximilian Lübke\IEEEauthorrefmark{5},
Jacob Kochems\IEEEauthorrefmark{1}, Stefan Köpsell\IEEEauthorrefmark{3},  Anosch Aziz-Koch\IEEEauthorrefmark{4},\\  Julia Beuster\IEEEauthorrefmark{2}, Oliver Blume\IEEEauthorrefmark{7}, Norman Franchi\IEEEauthorrefmark{5}, Reiner Thomä\IEEEauthorrefmark{2}, S{\l}awomir Sta\'{n}czak\IEEEauthorrefmark{6} and Hans D. Schotten\IEEEauthorrefmark{1} } 
\IEEEauthorblockA{\IEEEauthorrefmark{1}%
WiCoN,
University of Kaiserslautern-Landau (RPTU),
Germany, \{firstname.lastname\}@rptu.de
}
\IEEEauthorblockA{\IEEEauthorrefmark{2}%
TU Ilmenau,
Germany, \{firstname.lastname\}@tu-ilmenau.de\\
}
\IEEEauthorblockA{\IEEEauthorrefmark{3}%
Barkhausen Institute,
Dresden, Germany, \{firstname.lastname\}@barkhauseninstitut.org\\
}

\IEEEauthorblockA{\IEEEauthorrefmark{4}%
NXP Semiconductors Germany GmbH, 22529 Hamburg, Germany, williamhermann.melitiwa@nxp.com\\
}

\IEEEauthorblockA{\IEEEauthorrefmark{5}%
ESCS, Friedrich-Alexander-Universität Erlangen-Nürnberg, Germany, \{firstname.lastname\}@fau.de\\
}

\IEEEauthorblockA{\IEEEauthorrefmark{6}%
Fraunhofer Heinrich Hertz Institute,  Germany, \{firstname.lastname\}@hhi.fraunhofer.de\\
}

\IEEEauthorblockA{\IEEEauthorrefmark{7}%
 Robert Bosch GmbH, Advanced Engineering, 71272 Renningen, Germany, oliver.blume@de.bosch.com
}
}
%Einverständnis Projektpartner:
% RPTU: OK
% FAU: OK
% Ulm: OK
% Hensold: OK
% NXP: OK 
% BI: OK 
% Bosch: ?
% Denso: OK ?
% TU Ilmenau: OK 
% CIS: OK
% HHI: OK 
% IMST: OK
% MLE: ?
% Merantix: ?
% AeroDCS: OK

\maketitle

\begin{abstract}
Sensing plays a crucial role in autonomous and assisted vehicular driving, as well as in the operation of autonomous drones. 
The traditional segregation of communication and onboard sensing systems in mobility applications is due to be merged using \gls{JCAS} in the development of the 6G mobile radio standard.
The integration of \gls{JCAS} functions into the future road traffic landscape introduces novel challenges for the design of the 6G system architecture.
Special emphasis will be placed on facilitating direct communication between road users and aerial drones. 
In various mobility scenarios, diverse levels of integration will be explored, ranging from leveraging communication capabilities to coordinate different radars to achieving deep integration through a unified waveform. 
In this paper, we have identified use cases and derive five higher-level \glspl{TC}.
Technical and functional requirements for the 6G system architecture for a device-oriented \gls{JCAS} approach will be extracted from the \glspl{TC} and used to conceptualize the architectural views.
%echnical and functional requirements for the 6G system architecture, tailored for a device-oriented \gls{JCAS} approach, are extrapolated from these Tech Cases and utilized to conceptualize different architectural views.
%Sensing is an essential function for autonomous and assisted vehicular driving and for autonomous drones. Onboard sensing can be enhanced by communication between vehicles or even by a combination of sensing and communication. The use of \gls{JCAS} functions in the road traffic of the future poses new challenges for the development of the 6G system architecture. Special attention will be given to direct communication between road users and aerial drones. In relevant mobility use cases, different levels of integration will be considered, ranging from using communication capabilities to coordinate different radars to deep integration using a single waveform. From this collection of use cases, 5 higher-level tech cases are derived. Technical and functional requirements of the 6G system architecture for a device-oriented \gls{JCAS} approach are extracted from the tech cases, which are used to conceptualize the architectural views.
\end{abstract}
% \vspace{0.5 cm}
\begin{IEEEkeywords}
6G, Architecture, \gls{JCAS}, V2X, U2X 
\end{IEEEkeywords}

\section{Introduction}
In an era marked by the rapid advancement of communication and radar technologies, research with a prominent focus on the convergence of these two domains has gained significant momentum~\cite{Wymeersch2021, Wild2021, Qian2022}. This convergence, referred to as \gls{JCAS}, stands as a key technological enabler for 6G systems. It holds the potential for efficient use of scarce spectrum resources, cost effectiveness, reduced energy consumption by enabling dual functionality for various applications, such as \gls{ITS} \cite{Thoma2019}, smart factories, and more. Depending on the applications, different \glspl{CL} have been defined in~\cite{stanczak20226g}:

\begin{enumerate}
    \item Separation (\textbf{\gls{CL}1}): Communication and radar sensing operate in stand-alone mode with complete disregard of each other.
    \item Coexistence (\textbf{\gls{CL}2}): Sensing benefits from the hardware and software developments of 6G communication, e.g., via scheduling for interference mitigation.
    \item Cooperation (\textbf{\gls{CL}3}): Sensing enabled by wireless connectivity, as in collaborative radar networks, or communication enhanced by contextual information from sensing and/or localization.
    \item Integration (\textbf{\gls{CL}4}): Reuse of communication signals, hardware, and infrastructure for sensing (partial integration), or joint design for time- and frequency-concurrent operation of sensing and communication (deep integration).
\end{enumerate}
\,\,\,\,Automobiles and drones/\gls{UAV} are prime candidates for the inclusion of \gls{JCAS} technology. Their existing communication and radar systems present a wealth of untapped potential. Since release 14, 3GPP has standardized \gls{V2X} protocols for direct wireless communication from vehicles to other vehicles and infrastructure components like \glspl{RSU}. This so-called sidelink reuses the same frequency bands as cellular links. In the context of 5G, we can thus compare sub-6 GHz and \gls{mmW} sidelink, where the latter entails higher data rates and spatial precision but also higher attenuation \cite{Wymeersch2021}.
However, integrating these technologies comes with its own set of challenges. These challenges arise from the dynamic nature of these applications and their environments, stringent requirements for security and privacy of sensing data, the choice of integration levels and degrees of coordination, the quest for optimal performance, the development of specialized hardware such as antenna concepts, and the aggregation, fusion, and analysis of complex, often distributed, sensing data. Designing a system architecture to address these challenges will be essential in unlocking the full potential of \gls{JCAS} systems in the evolving landscape of communication and radar technologies.

To investigate the possibility of incorporating \gls{JCAS} in 6G networks, a sole description of a monolithic architecture cannot meet all requirements, and traceability cannot be ensured in this way \cite{karrenbauer2019future,ludwig2018a5g}. Therefore, the overall architecture is regarded concerning various aspects of the system, which involves identification of potential future applications in the automotive and drone sectors and documenting the system requirements associated with them.
The rest of the paper is outlined as follows. Section II summarizes the related work found in literature. Section III identifies the Tech Cases and describes their system requirements. Section IV describes the different architecture views. Lastly, Section V concludes the paper.

\section{Related Work}
In Release 17, 3GPP has described positioning services to support verticals and applications with positioning accuracies better than 10 meters \cite{TS22.261}. Currently, the positioning functions are located at the \gls{5GC} \cite{TS23.273}, however to fulfill future application requirements, these positioning functions are envisioned to be deployed in 6G networks at the edge \cite{Mogyorósi2022}.
Other functionalities and technological enablers for 6G, such as \gls{JCAS}, \gls{DT}, \gls{AI}, cloud computing are also needed to fulfill the futuristic mobility and industrial applications \cite{fettweis2021joint,EU6G2023}. 
Various uses cases and their requirements for \gls{JCAS} have been described by 3GPP \cite{TR22.837} and by the Milestone Document of the KOMSENS-6G project \cite{KOMSENS}.
A large number of publications explain the profitable advantages of \gls{mmW} and sub-THz bands in 6G for implementing JCAS at base stations.
In \cite{Wild2021}, essential system design aspects for the deep integration of cellular-based \gls{JCAS} (\gls{CL}4) are described, with a special focus on the choice of waveforms for communication-centric radar sensing and the integration into the existing 5G NR. Another example is \cite{LMP}, which describes a novel transceiver architecture and frame structure for \gls{CL}2 and \gls{CL}4 \gls{JCAS} at base stations and a processing framework. A processing framework relying on \gls{mmW} massive multiple-input multiple-output (MIMO) systems were presented by \cite{GaoZhen}.
Some publications describe smaller blocks of a \gls{JCAS} system architecture, such as  \cite{KrauseA} which proposes a machine learning (ML)-based approach that combines knowledge from a distributed sensor network and knowledge on obstacles to create an radio environment map (REM) in industrial production (\gls{CL}3). Another detailed \gls{JCAS} \gls{CL}3 architecture description for an efficient transmission and secure sharing architecture of sensing data based on federated learning has been proposed by \cite{MuJunsheng} and \cite{Ouyang}.
\gls{JCAS} variants covered in the literature typically rely on communication infrastructure, such as base stations. In contrast to this, we foresee a device-centric development in the mobility context, where sensing and processing are performed in end devices with an important use of sidelink communication. To the best of our knowledge, no comprehensive 6G system architecture in the \gls{JCAS} for mobility context has been proposed.

\section{Relevant \glsentrylongpl{TC} and their Requirements} \label{sec:use_case_to_tech_case}
The industrial project ICAS4Mobility has identified 5 broad \glspl{TC} that take into account over 20 particular use cases for \gls{JCAS} in a mobility context. 
 A brief description of the \glspl{TC} and their requirements are provided below.

\subsection{Description of \glsentrylongpl{TC}}
\begin{enumerate}
    \item \textbf{\gls{TC}-1 - Sidelink Based Aerial Environment Sensing}:
    This \gls{TC} applies \gls{CL}4 \gls{JCAS} over sidelink for a single drone. 
    The drone can plot its course and navigate through an unfamiliar environment without relying on base station coverage. 
    The fusion of inertial, radar and optical data improves the accuracy and reliability of sensing, and enhances the ability to identify and track objects.
    
    \item \textbf{\gls{TC}-2 - Collaborative Radar for Terrain and Building Mapping:}
    In this \gls{TC}, all sensing requirements of the \gls{TC}-1 are extended to a collaborating swarm of drones. 
    Raw or pre-processed data is communicated to a (mobile) control station on the ground. 
    A primary drone can be used as a multi-static sensing and communication platform to the ground station. 
    Central drone management via sidelink communication (\gls{CL}3) facilitates the cooperative swarm operation.

    \item \textbf{\gls{TC}-3 - Identification and Mapping of Non-Cooperative Radio Emitters:}
    The aim of this \gls{TC} is to apply \gls{CL}3 to utilize the radio interface for the purpose of detecting and localizing foreign radio emitters by fusing observations made by distributed sensing nodes. This can include both interfering and non-connected transceivers. Since these emitters are external (read: \emph{non-cooperative}), this \gls{TC} is clearly distinguished from the emitter localization approach taken by the 3GPP that only localizes \emph{cooperative} transmitters. It is useful to note that the system possesses no knowledge about the radio transmissions to be localized, i.e. the carrier frequency, bandwidth, modulation format and access scheme are unknown. 
    
    \item \textbf{\gls{TC}-4 - Vehicular-based Communication-centric Sensing:}
    This \gls{TC} deploys CL4 \gls{JCAS} over the sidelink to scan the wayside and roadside, in addition to conventional radar in vehicles. The car can provide radar measurement results to other vehicles or to a centralized SLAM server for a dynamically updated map of parking lots and of objects relevant for the maneuvers of vehicles. Due to the limited bandwidth of the 6GHz sidelink, the concept of \gls{SAR} could be used for obtaining highly precise angle estimations of objects along the road.

    \item \textbf{\gls{TC}-5 - Cooperative Vehicular Radar Sensing:}
    The \gls{TC} of cooperative vehicular radar sensing envisions a multi-static radar system for object detection on the road. This requires \gls{CL}3 communication to achieve different collaboration levels that improve the sensing quality compared to a stand-alone radar. The \gls{TC} focuses on the exchange of processed data between radar nodes and the utilization of 77 GHz radar hardware, which features mono- or bi-static sensing and exchange of point clouds.
\end{enumerate}

\subsection{System Requirements}

In this section, we state the technical and functional requirements of the previously presented \glspl{TC}.
The later used requirement verbs represent different classes of requirements: \textit{shall} corresponds to absolute mandatory requirements, \textit{should} refers to recommended requirements and \textit{may} addresses optional ones.

\begin{enumerate}
    \item \textbf{Management:}
    The system should accommodate the different precision requirements for time synchronization between sensor nodes as required by each integration level (\textbf{R1}). The system shall also facilitate smart data augmentation management, allowing for a secure balance between individual node data usage and collective information from multiple nodes (\textbf{R2}). Additionally, the system shall ensure data redundancy in compliance with safety, privacy, and security standards, specifically following automotive norms such as ISO 26262 (\textbf{R3}).
    \item \textbf{Infrastructure:}
    The system shall support on-the-edge processing of data. This processing may occur on the same node as the data is generated from, or on nearby (sensing) nodes that possess the required computational resources (\textbf{R4}).
    The system may support a cellular link (\textbf{R5}). 
    The system shall be capable of operating a 6\,GHz antenna with a pattern that covers the lateral direction of the vehicle (\textbf{R6}).%adjustable directional characteristic (corner radar)  % R57

    \item \textbf{JCAS:}

    The system shall support a communication protocol to mitigate interference between sensing nodes, either via sidelink communication, or via communication with a central base station (\textbf{R7}). A sensing entity should be able to be tuned to multiple carrier frequencies and operate at an adaptable bandwidth to reach \gls{CL}2 (\textbf{R8}).

    The system shall be capable to operate the radar functionality in a cooperative way where only pre-processed data (point clouds or object lists) but no raw data is shared (\gls{CL}3) (\textbf{R9}).
    The system shall integrate sensing capabilities into a communication-capable platform (\gls{CL}4) to reuse communication signals, hardware and infrastructure for sensing (\textbf{R10}).
    
    \item \textbf{Communication:}

    All entities in the system shall support the transmission of sensor data labeled with additional meta-data denoting the properties of the source sensor, as well as support dynamic radio resource allocation based on current sensing requirements (\textbf{R11}).
    The system shall provide sufficient bandwidth to exchange point clouds and images with neighboring drones or vehicles (\textbf{R12}).
    The system may share sensing results to a centralized server using cellular link or via sidelink to surrounding vehicles (\textbf{R13}).

    \item \textbf{Sensing:}

    The system shall support fault-tolerant sensing functionalities within a node or between cooperative sensing nodes and provide full-duplex mono-static sensing capability of signal processing (\textbf{R14}).
    The sensing node shall be able to detect radio signals with unknown modulation schemes and provide a high availability of position measurements when sensing is used as a replacement of \gls{GPS} positioning (\textbf{R15}).
    The sensing radar shall be capable of \gls{SAR} signal processing to detect objects/free space from the ego-vehicle perspective and may support the ability to be tuned to multiple carrier frequencies and operate at an adaptable bandwidth (\textbf{R16}).
    
    \item \textbf{Processing:}
    % Data Fusion
    The system shall support fusion of both physical and virtual sensor data from various sensor nodes such as cameras, including interference dependencies, and detection of signals through collaborative radars (\textbf{R17}). The system shall enable  distributed data fusion among cooperating sensing nodes (\textbf{R18}).

    The system should efficiently classify detected emitters based on radar illumination and sensor observations (\textbf{R19}). It shall reliably detect and classify objects within defined size and range criteria, while also detecting and localizing non-cooperative electromagnetic emitters, such as radars and communication devices (\textbf{R20}).

    The system shall localize detected emitters, signals within a node through areal environment sensing, and radar signals between cooperative sensing nodes (\textbf{R21}). It shall provide mapping based on areal environment sensing and receive high-resolution maps for \gls{SLAM} (\textbf{R22}). Additionally, the system shall perform in-vehicle \gls{SLAM} processing to dynamically update maps and enhance the precision of the ego vehicle and its trajectory (\textbf{R23}).
    
    \item \textbf{Privacy and Security:}

    The system shall provide (long-term) integrity, confidentiality, authenticity, and privacy-preserving data processing (\textbf{R24}). It shall be compliant with the European \gls{GDPR} (\textbf{R25}), shall provide a \gls{PKI}, and shall ensure use-case-aware and long-term security and privacy (\textbf{R26}).
        
\end{enumerate}

\section{Architecture Views}
From the investigation of the \glspl{TC} presented in the previous section, we see that a system architecture capable of supporting the associated requirements must be designed with flexibility, robustness and security in mind. Therefore, we propose various architectural perspectives, including Node, Functional, Data Flow, and Privacy and Security, which are described below.
\subsection{Node-View}
\begin{figure*}[!ht]
\centering
\resizebox{\textwidth}{!}{
\includegraphics[]{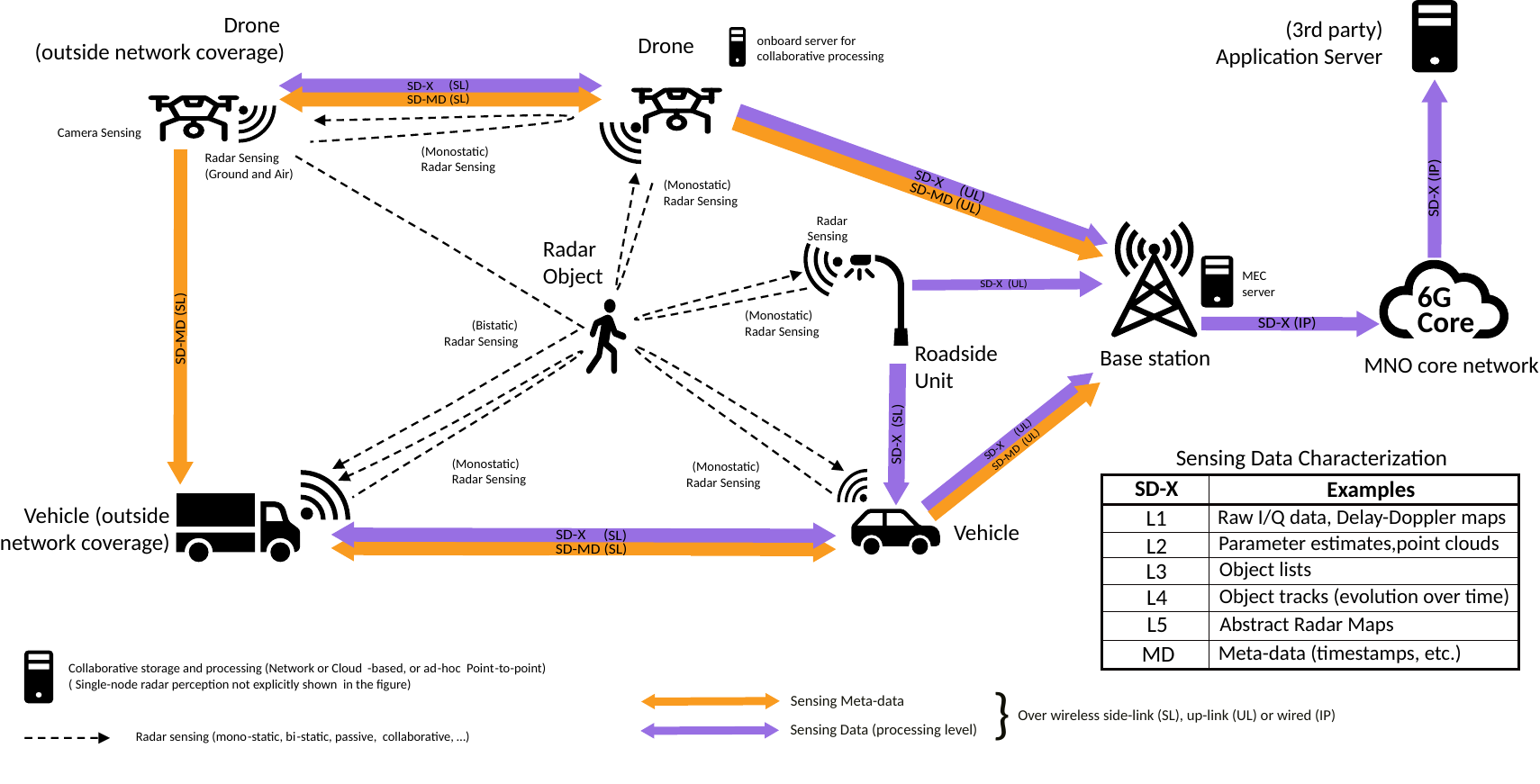}
}
%\label{fig:node_view}
\caption{This Figure illustrates the combined Node and Data-flow views of the proposed architecture. The different types of sensing nodes are illustrated with dotted lines signifying the radar transmissions. The thick arrows signify the flow of sensing data long with relevant meta-data.}
\vspace{-0.3cm}
\label{fig:node_view}
\end{figure*}

The node view on the architecture, which is depicted in Fig.~\ref{fig:node_view}, represents the physical components of the architecture. The proposed architecture partly combines existing systems into a system of systems interacting by using communication links.

In the following, a detailed overview of the components of the node architecture is provided.

\begin{enumerate}
    \item \textbf{Vehicle:} Vehicles play a pivotal role as a connected and intelligent entity within the broader transportation ecosystem (\textbf{R10}).
    The vehicle's integration of radar sensing, 6G mobile communication, and sidelink capabilities transforms it into an intelligent, data-sharing entity that enhances both its own operational efficiency and the overall safety and efficiency of the transportation system (\textbf{R3}).
    %The vehicle uses radar sensors to constantly scan its surroundings, detecting and tracking nearby objects, vehicles, pedestrians, and obstacles.
    %By analyzing the radar data, the vehicle gains a real-time understanding of its environment, allowing it to make informed decisions to ensure safety.
    %Leveraging 6G connectivity, the vehicle establishes high-speed, ultra-reliable communication links with both the RSU and other nearby vehicles.
    %This enables the vehicle to exchange a vast amount of data rapidly, including traffic conditions, road hazards, and navigation updates.
    %The vehicle engages in sidelink communication to directly communicate with neighboring vehicles and the RSU without relying on a centralized network.
    %This direct communication enables the vehicle to share critical information with other vehicles, such as its position, speed, and intended maneuvers.
    %It can also receive warnings and messages from other vehicles and the infrastructure.
   
    \item \textbf{Drone:} Similarly to vehicles, drones can be equipped with various sensor types, including cameras, \gls{LiDAR}, and radar sensors (\textbf{R14}, \textbf{R16}).
    %They can fly over roadways and gather data about traffic conditions, road quality, and incidents.
    The acquired data can be transmitted to the central system or nearby \glspl{RSU} or directly between them or other vehicles via communication links (\textbf{R13}).
    %Drones can assist in traffic management by providing real-time traffic updates from an aerial perspective.
    %They can help identify congestion, accidents, or roadblocks and communicate this information to RSUs and vehicles, allowing for faster responses and detours.
    %Drones can be rapidly deployed to provide situational awareness during emergencies.
    %For example, in the case of a traffic accident or natural disaster, drones can capture high-resolution images and videos, helping authorities assess the situation and plan their response.
    Drones can serve as controllable communication relays in areas with poor network coverage (\textbf{R12}) by establishing temporary communication links.
   
    \item \textbf{\gls{RSU}:} \glspl{RSU} are crucial components in modern \gls{ITS} designed to enhance road safety and traffic efficiency \cite{Thoma2019}.
    In the scope of this architecture, the \gls{RSU} is equipped with radar sensing technology and utilizes both %6G and sidelink communication.
    traditional mobile and sidelink communications (\textbf{R5}, \textbf{R13}).
    %The RSU uses radar technology to sense objects and vehicles in its vicinity.
    %With this it is able to determine the presence, speed, distance, and trajectory of objects.
    %This enables the RSU to gather real-time data about the traffic flow, detect obstacles, and monitor vehicle movements.
    %The RSU leverages 6G connectivity to establish high-speed, low-latency wireless communication with nearby vehicles, infrastructure, and central traffic management systems.
    %Sidelink communication involves direct communication between nearby devices or units without relying solely on a centralized network.
    In this context, the \gls{RSU} uses sidelink communication to establish direct connections with nearby vehicles equipped with compatible communication systems.
    %This enables the RSU to exchange critical information, such as collision warnings or traffic updates, directly with vehicles in its proximity.
    
    \item \textbf{\gls{MEC}:}
    %Edge computing refers to the practice of processing data closer to the data source, at the "edge" of the network, rather than sending all data to centralized cloud servers.
    Edge computing resources are co-located with base stations and enable real-time data processing \cite{Xiao2020}, low-latency decision-making, and improved system resilience (\textbf{R2}, \textbf{R4}, \textbf{R17}, \textbf{R22}).
    %If a group of vehicles or drones are out of range of the nearest base station, a mobile co-located solution can provide computing resources for data processing, while the vehicle group is detached from the base station.
    If a group of vehicles or drones is outside the range of the nearest base station, a mobile co-located solution can provide computing resources for data processing while the group of vehicles is separated from the base station.
    %We will refer to this concept as “Agile Mobile Edge Computing”.
    %The low-latency nature of edge computing ensures that the time between data capture and data processing is minimized.
    %This is vital for safety-critical applications.
    %Furthermore, edge computing can filter and prioritize data before transmitting it to higher-level systems or the cloud.
    %This minimizes the amount of data that needs to be sent over the network, optimizing bandwidth usage and reducing potential congestion.
    %Finally, edge computing enhances system resilience.
    Even if the central cloud server experiences a disruption, edge devices can continue to operate and communicate with each other, ensuring that critical functions are still available (\textbf{R14}).
    
    \item \textbf{Radar Object:}
    Radar objects are physical objects like vehicles, pedestrians, or obstacles, that reflect the signals emitted by radar sensors such that the system can detect and continuously track the objects' positions, speeds, and trajectories (\textbf{R9}).
    Each radar object is represented as a data entity with attributes such as position, velocity, and possibly additional characteristics like size or reflectivity. These attributes are continuously updated as the radar sensor receives new data about the radar objects (\textbf{R18}, \textbf{R21}, \textbf{R23}).
    %The information about radar objects can be shared through 6G and sidelink communication with other vehicles and infrastructure elements like RSUs.
    %This communication enables nearby vehicles to be aware of the presence and behavior of radar-detected objects, facilitating cooperative safety measures and traffic coordination.
    
    %\subsubsection{6G Communication}
    %The 6G communication system provides advanced and high-performance wireless communication capabilities.
    %6G communication offers exceptionally high data transmission speeds, surpassing the capabilities of previous generations.
    %It supports the real-time sharing of sensor data, traffic information, and updates crucial for intelligent transportation.
    %6G technology reduces communication latency to near-instantaneous levels.
    %This low latency is essential for time-sensitive applications such as autonomous driving and safety-critical communications.
    %6G is expected to support a massive number of connected devices per square kilometer.
    %This is particularly relevant in an intelligent transportation context where numerous vehicles, roadside units, and other sensors need to communicate simultaneously.
    %It facilitates the seamless integration of all these elements into a cohesive system.
\end{enumerate}

Furthermore, sensing nodes can operate individually or in a co-operative manner by forming groups of collaborative nodes. In a co-operative setups, one or more node(s) are appointed the \emph{leader} and facilitate communication with the infrastructure for all nodes within the group.

\subsection{Functional-View}
\begin{figure}[t]
\centering
\resizebox{\columnwidth}{!}{
\includegraphics[]{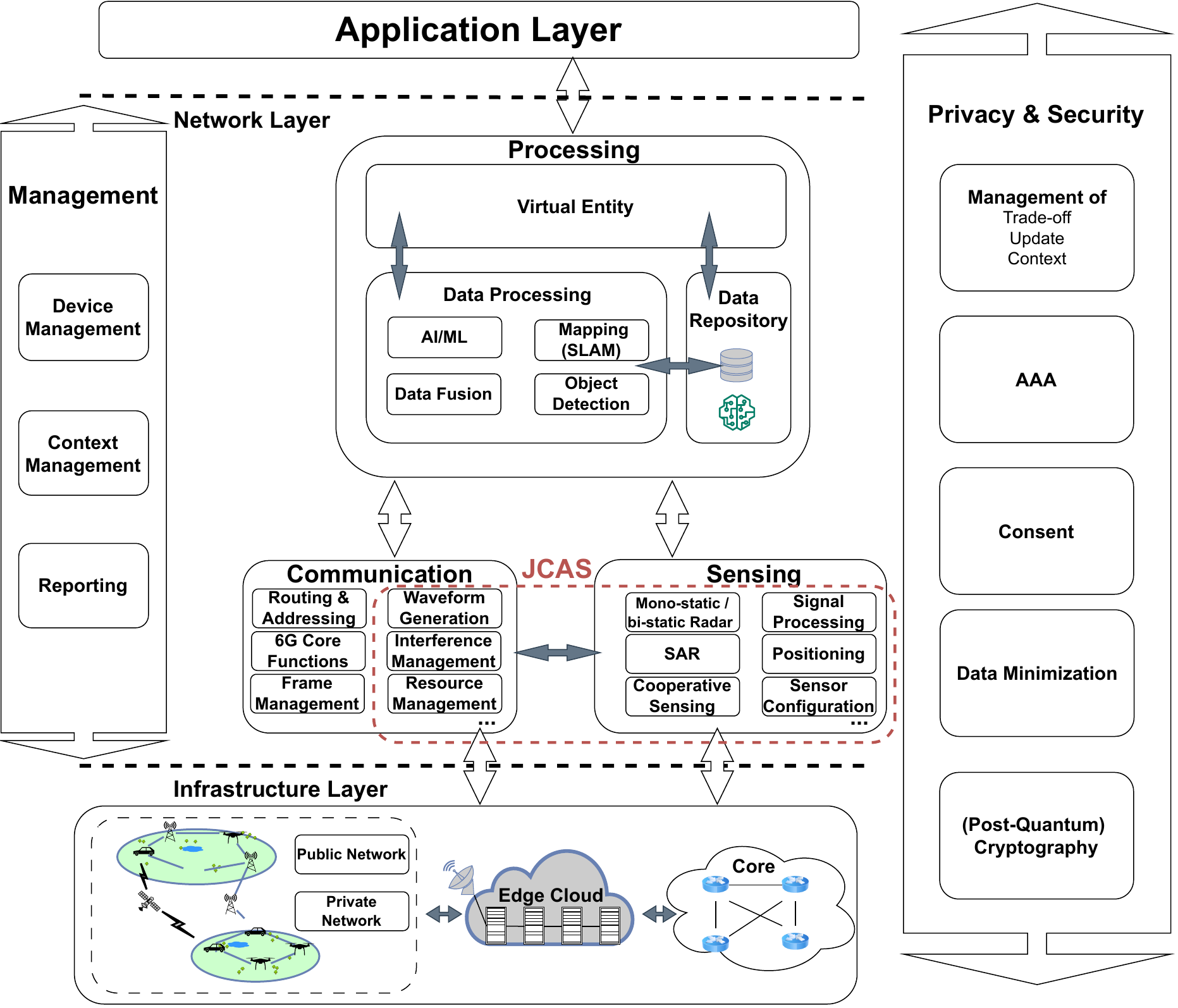}
}
\caption{The depicted functional view image is divided into the infrastructure layer, network layer and application layer. The functional view consists of the blocks: Management, Processing, Communication, Sensing, Privacy and Security, as well the \gls{JCAS} block, marked in red.}
\label{fig:functional_view}
\vspace{-0.3cm}
\end{figure}

The functional view (see \Cref{fig:functional_view}) aims to describe the required system blocks for device-driven \gls{JCAS} functionalities.
It interconnects the network layer, the application layer and the infrastructure layer as outlined in the node-view. Furthermore, these layers interact with the privacy and security block, detailed in the Privacy and Security subsection. The requirements from our \glspl{TC} are fulfilled by the following five main functional blocks, along with the interaction between the layers.
\begin{enumerate}
    \item \textbf{Management:} 
    Management block is responsible for \gls{E2E} wireless network, including assignment, control, management and reporting of devices, services and infrastructure alike. For example, this block is responsible for time synchronization (\textbf{R1}), smart data augmentation (\textbf{R2}), data redundancy amongst sensor nodes (\textbf{R3}). Moreover, it also ensures that all devices included in the system comply to local and international laws and standards (\textbf{R3}).
    
    \item \textbf{Communication:}
    Communication block is responsible for transmission of messages between sensing nodes and wireless network infrastructure to ensure application-specific latency and \gls{QoS} requirements. This includes, but is not limited to, routing of messages, interference management, resource management, waveform generation. Depending on the message type (\glspl{CAM}, point clouds, meta data, object lists) and requirements of the sensing nodes, this block will dynamically allocate bandwidth, time slots and frequencies to enable different \gls{V2X}/\gls{U2X} communication modes (\textbf{R12}, \textbf{R13}).
    
    \item \textbf{Sensing:}
    The sensing block monitors the physical environment via a variety of sensors like \gls{LiDAR}, cameras, radars.
    The radar sensing subsystem is capable of working with mono-/bi-/multi-static configurations (\textbf{R14}).
    Depending on the application the operational parameters like frequency, bandwidth and modulation schemes are dynamically adjusted (\textbf{R14}).
    Moreover, it includes signal processing techniques for \gls{SAR} imaging via a sidelink channel (\textbf{R15}).
    
    \item \textbf{Processing:}
    This block includes three closely integrated sub-blocks, namely data repository, data processing and virtual entity. 
    Data repository is responsible for the collection of real-time data from sensing nodes and infrastructure for storage in heterogeneous databases.
    Data Processing refers to \gls{AI} and statistical models leveraged to perform analysis, prediction, emulation, diagnosis, localization, mapping (\gls{SLAM}), classification, and fusion for various application scenarios by using stored data from the data repository.
    Depending on the application and scenario requirements, different \gls{AI} and statistical models can be instantiated, trained, updated and stored for future use (\textbf{R17}, \textbf{R18}, \textbf{R19}, \textbf{R20}, \textbf{R21}, \textbf{R22}, \textbf{R23}).
    Virtual entity is an added functionality that provides a safe virtual environment for simulations and could be evolved into a  \gls{DT}.
    
    \item \textbf{\gls{JCAS}:}
    %The \gls{JCAS} functionality is the convergence of the two subsets of the communication and sensing functionalities respectively.
    \gls{JCAS} enables functionality such as cooperative sensing, interference management, sidelink \gls{SAR}, and dual usage of waveforms for both communication and sensing. This functionality allows for the use of both communication-centric sensing and sensing-centric communication at different integration levels by reusing hardware, signals, and spectrum resources, depending on the application (\textbf{R7}, \textbf{R8}, \textbf{R9}, \textbf{R10}).
\end{enumerate}

\subsection{Data-Flow}
The Data-Flow view describes the generation and transmission of data at a sensing node that is either processed locally or transmitted to other entities within the system for further processing or data fusion. This architecture view characterizes the different kinds of signals and data that should be exchanged between different entities in the node view to enable \gls{JCAS} functionality. Our architecture proposes three different kinds of data between entities, namely:
\begin{enumerate}
    \item \textbf{Sensing Data:} 
Sensing data includes radar observations made by a sensor node along with relevant meta-data required to process this data at a fusion center or within a collaborating node (\textbf{R9}, \textbf{R14}, \textbf{R18}). Due to the different kinds of data generated and shared for each \gls{TC} in \Cref{sec:use_case_to_tech_case}, we define a generalized sensing data-flow labelled \textbf{SD-X} between nodes in \Cref{fig:node_view}. This abstract description is specified based on the level of pre-processing the data has undergone before further transmission. Examples for the type of data described by each level are given in the table in \Cref{fig:node_view}.

Every sensing data-flow is accompanied in \Cref{fig:node_view} by meta-data (labelled \textbf{SD-MD}) that contains information required to further utilize the sensing data, e.g. time-stamps or sensing node configuration (\textbf{R1}, \textbf{R2}, \textbf{R16}).

\item \textbf{Control Signals:}
Control signals contain information that configure the sensor nodes and originate either from the 6G infrastructure or from leader nodes in a collaborative sensing setup (\textbf{R7}, \textbf{R10}). They may include parameters that define the required radar resolution or numerology used in the \gls{PHY} layer.

\item \textbf{Synchronization Signals:}
These signals enable synchronization between collaborating radar nodes, and are further classified based on the synchronization precision required between nodes for the specific \gls{TC} (\textbf{R1}). The precision required further depends on the functionality they enable:

\textbf{Class 1}: Time-stamping of data for further processing

\textbf{Class 2}: MAC Frame synchronization

\textbf{Class 3}: TX-RX Clock synchronization
\end{enumerate}

\subsection{Privacy and Security}

The privacy and security architecture view describes the principles that are applied to resist the threats opposed by vehicles and drones, equipped with strong sensing capabilities, operating in public areas.
This threat is mostly raised by the fact that modern radar technologies can biometrically identify people, \cite{pfeuffer2019behavioural,kyoso2001development,biel2001ecg}, and even recognize (through-wall) human movement and behavior \cite{huang2021indoor,qi2016detection}.
Furthermore, the architecture addresses security (and thus safety) risks. Those are, among others, manipulations regarding the collection, transmission, or processing of sensing data. 
Or through directly attacking the system using, for instance, network attacks, denial of service attacks, or quantum computer attacks.
We propose the following components and principles.

\begin{figure}[!ht]
\centering
\resizebox{0.5\textwidth}{!}{
\includegraphics{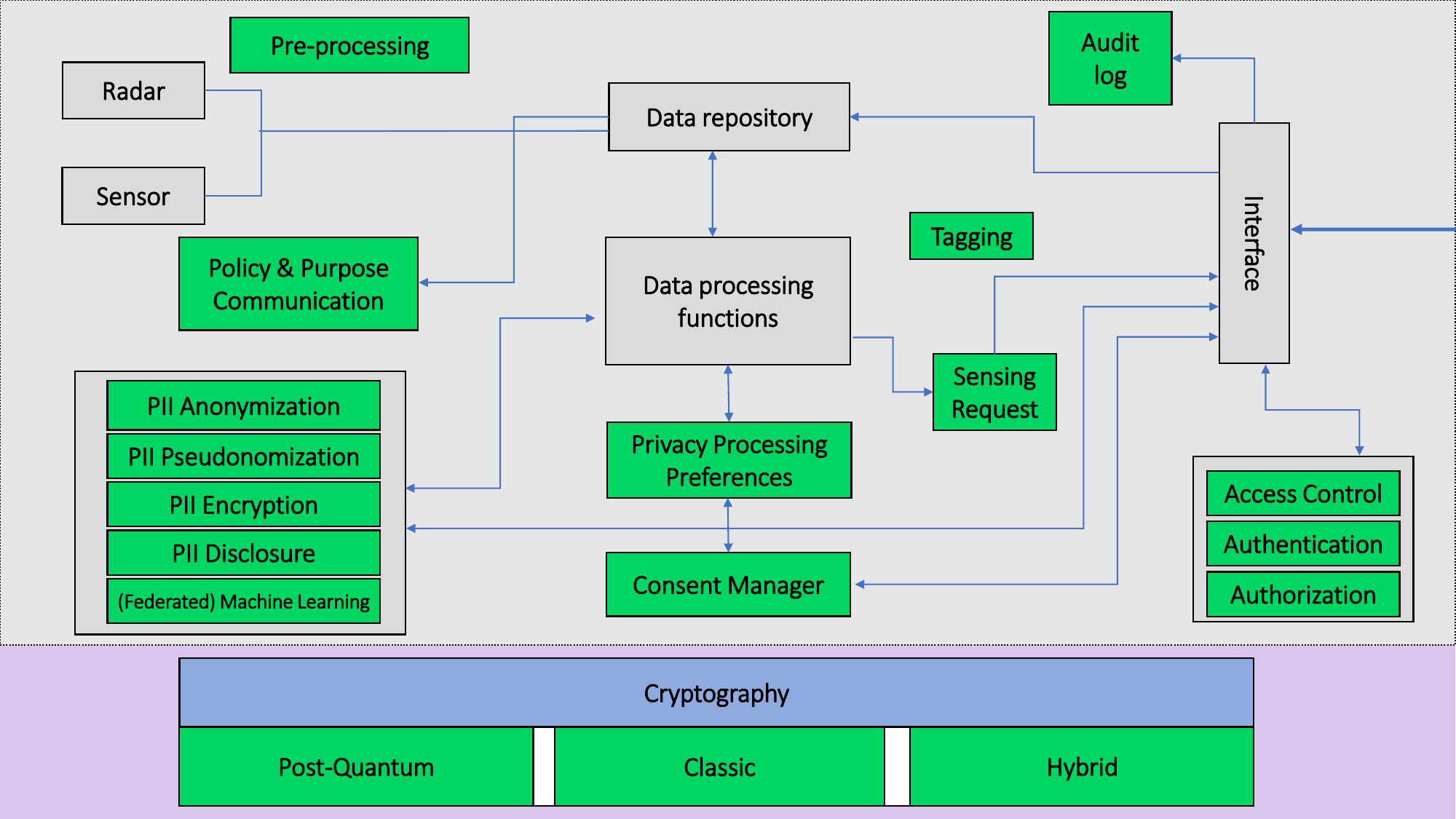}
}
\caption{This Figure illustrates the security and privacy architecture. The green boxes are the applied principles, aligned with the grey boxes taken from the functional view in Figure 2.}
\vspace{-0.3cm}
\end{figure}

\begin{enumerate}
    \item \textbf{Privacy Controls:}
    Meeting the requirements of data security and data protection necessitates the implementation of a set of security and privacy (S\&P) controls. 
    These controls ensure that data is processed securely and in compliance with \gls{GDPR} regulations (\textbf{R25}). 

    \item \textbf{Access Control:}
    In our architecture (following the 3GPP service-based architecture design), a single interface is foreseen for access to sensor data, facilitating the management of data requests. 
    The policy-driven decisions regarding the access to sensor data as well as the S\&P controls to be applied are contingent on factors such as data granularity, sensitivity, intended use, and the requesting entity (\textbf{R25}).
    
    \item \textbf{(Post-Quantum) Cryptography:}
    Therefore, an important first step involves the establishment of identities and corresponding access control mechanisms. Those are fortified with cryptographic security measures against attacks issued both by classical and quantum adversaries, as functionalities that provide authentication and authenticity (\textbf{R24, R26}).
    
    \item \textbf{Tranparency:}
    One core principle of \gls{GDPR} is transparency, ensuring that the affected human knows the collecting entity and the purpose of collection.
    In the context of \gls{JCAS}, transparency can be improved by utilizing the communication means of \gls{JCAS} to communicate the transparency information, for instance via broadcasting the information (\textbf{R25}).
    
    \item \textbf{Consent:}
    Consent is one way to allow the processing of personal data, which is forbidden by \gls{GDPR}. Its applicability for \gls{JCAS} needs to be analyzed from legal and technical perspectives.
    To foster research, we propose a dynamic system where users' privacy policies are communicated in real-time, supporting users to enforce their data-sharing preferences. This raises several questions such as feasibility, how to implement such controls, and are those solutions in line with \gls{GDPR} (\textbf{R25}).
    %
    %a) Is it feasible (from a technical and usability perspective) to enable users to constantly and in real-time communicate their privacy preferences? b) How to implement related S\&P controls that enforce these policies on the sensing side? c) Is the resulting solution appropriated from a regulatory and legal standpoint for compliance with GDP?
    
    \item \textbf{Data Minimization:}
    Our architecture puts a strong emphasis on data protection principles, with a focus on data minimization. 
    This minimization is achieved by processing data as close to the sensor as possible, not only to conserve bandwidth and technical resources but also to enhance privacy.
    Data minimization can be achieved by applying advanced technologies such as machine learning, federated learning, or the usage of \gls{ORAN}. 
    In cases where on-device data processing is not feasible, alternative techniques such as data obfuscation, generalization, and noise injection should be considered (\textbf{R24, R25, R26}).
    
    \item \textbf{Context Awareness:}
    Lastly, ensuring data privacy requires a profound understanding of context. 
    This entails making informed decisions about which data is pre-processed and disclosed to whom, in accordance with users' privacy preferences. (\textbf{R26})
    Achieving this context awareness is a multifaceted challenge that needs comprehensive investigation.
\end{enumerate}

\section{Conclusion}
This paper propose a general system architecture for the 6G network in the context of future automotive and aerial drone mobility. We identify different \glspl{TC} and characterize their system requirements. We also provide different architectural views, namely node-view, functional-view, data-flow, and privacy and security view, and defined all the individual entities belonging to each view. Furthermore, the system requirements are numbered and traced through the different architecture views to highlight that the different views not only conform to the requirements of the \glspl{TC}, but also provide enough flexibility to scale and add other functionalities to the system.

\section*{Acknowledgement}
This work has been supported by the Federal Ministry of Education and Research of the Federal Republic of Germany as part of the 6G-ICAS4Mobility project (16KISK229K). The authors alone are responsible for the content of the paper. Additionally we would like to thank our colleague Dieter Novotny (\textit{aeroDCS}) for his valuable insights. ChatGPT 3.5 has been used to improve the grammar of this paper.

\balance
\bibliographystyle{IEEEtran}
\bibliography{references}

\end{document}